\documentclass[conference]{IEEEtran}
\IEEEoverridecommandlockouts
\usepackage[utf8]{inputenc}
\usepackage{cite}
\usepackage{amssymb}
\usepackage{algorithmic}
\usepackage{graphicx}
\usepackage{textcomp}
\usepackage{xcolor}
\usepackage{tabularx}
\usepackage{multirow}
\usepackage{booktabs}
\usepackage{url}
\usepackage{pifont}
\usepackage{mdframed}
\usepackage{environ}
\usepackage{color, colortbl}
\usepackage{enumitem}

\NewEnviron{opus-box}[1][]{%
 \vspace{0.1cm}
    \begin{mdframed}[backgroundcolor=gray!25]
        \BODY
    \end{mdframed}
    \vspace{0.1cm}
}

\usepackage[pdftex,raiselinks=false,colorlinks=true,linkcolor=black,anchorcolor=black,citecolor=black,filecolor=black,urlcolor=black,linktocpage=true,bookmarksopen=true]{hyperref}

\newcommand{\gc}{\cellcolor{gray!25}}

\DeclareUnicodeCharacter{2212}{-}

\def\BibTeX{{\rm B\kern-.05em{\sc i\kern-.025em b}\kern-.08em
    T\kern-.1667em\lower.7ex\hbox{E}\kern-.125emX}}
    
\begin{document}

\title{On Relating Technical, Social Factors, and the Introduction of Bugs}

\author{
    \IEEEauthorblockN{
        Filipe Falcão\IEEEauthorrefmark{1}, Caio Barbosa\IEEEauthorrefmark{2}, Baldoino Fonseca\IEEEauthorrefmark{1}, Alessandro Garcia\IEEEauthorrefmark{2}, Márcio Ribeiro\IEEEauthorrefmark{1}, Rohit Gheyi\IEEEauthorrefmark{3}
    }
    \IEEEauthorblockA{
        \IEEEauthorrefmark{1}Computing Institute, Federal University of Alagoas, Maceió, Brazil\\
        \IEEEauthorrefmark{2}Pontifical Catholic University of Rio de Janeiro, Rio de Janeiro, Brazil\\
        \IEEEauthorrefmark{3}Federal University of Campina Grande, Campina Grande, Brazil\\
        \{filipebatista, baldoino, marcio\}@ic.ufal.br, \{csilva, afgarcia\}@inf.puc-rio.br, rohit@dsc.ufcg.edu.br
    }
}

\maketitle

\begin{abstract}
As collaborative coding environments make it easier to contribute to software projects, the number of developers involved in these projects keeps increasing. This increase makes it more difficult for code reviewers to deal with buggy contributions. Collaborative environments like GitHub provide a rich source of data on developers' contributions. Such data can be used to extract information about developers regarding technical (\textit{e.g.}, their experience) and social (\textit{e.g.}, their interactions) factors. Recent studies analyzed the influence of these factors on different activities of software development. However, there is still room for improvement on the relation between these factors and the introduction of bugs. We present a broader study, including 8 projects from different domains and 6,537 bug reports, on relating five technical, three social factors, and the introduction of bugs. The results indicate that technical and social factors can discriminate between buggy and clean commits. But, the technical factors are more determining than social ones. Particularly, the developers' habits of not following technical contribution norms and the developer's commit bugginess are associated with an increase on commit bugginess. On the other hand, project's establishment, ownership level of developers' commit, and social influence are related to a lower chance of introducing bugs.
\end{abstract}



\section{Introduction}
\label{sec:introduction}
Collaborative coding environments~\cite{tsay2014influence} like \textit{GitHub}, make it easier to contribute to software projects. However, these environments also make the evaluated contributions a challenging task to project managers and code reviewers~\cite{gousios2016work} for two reasons. First, developers with wide ranges of experience work simultaneously on the same projects. Second, developers can perform contributions along software development that may lead to the introduction of bugs~\cite{sliwerski2005changes}.

In spite of these challenges, collaborative code environments provide a rich source of data. This data can be explored to extract information on technical (such as developers’ experience) and social (such as interactions among developers) factors related to developers. Understanding the relation between those factors and the introduction of bugs may be useful for project managers and code reviewers. For example, let us consider the case a developer  with bug-related factors decides to perform a pull request in a project. Thus, code reviewers might want to double check this contribution, potentially avoiding the introduction of a bug.

Previous studies~\cite{tsay2014influence, gousios2014exploratory, gousios2016work, mockus2010organizational, eyolfson2011time, bernardi2012developers, bird2011don, posnett2013dual, guo2010characterizing} have used these factors to perform different analyses on developers' contributions. For instance, studies~\cite{tsay2014influence, gousios2014exploratory, gousios2016work} indicate that both technical and social factors impact the acceptance rate of contributions on GitHub. However, these studies did not analyze the influence of these factors on the introduction of bugs. Other studies~\cite{mockus2010organizational, eyolfson2011time, bernardi2012developers, bird2011don, posnett2013dual, guo2010characterizing} evaluated this influence only considering proprietary projects~\cite{mockus2010organizational, bird2011don, guo2010characterizing} or a reduced number of factors as well as characteristics to represent them~\cite{bird2011don,posnett2013dual}.

To improve the understanding on the relation between technical, social factors and introduction of bugs, we present a broader study involving five technical and three social factors. In particular, we analyze factors related to the developers' experience~\cite{eyolfson2011time, mockus2010organizational, rahman2011ownership, tufano2017empirical}, developers' habit to follow well-known technical contribution norms~\cite{tsay2014influence}, ownership~\cite{rahman2011ownership, bird2011don, thongtanunam2016revisiting}, nature of developers' changes~\cite{hattori2008nature}, bugginess of developer's commits~\cite{eyolfson2011time}, communication with the community of a project~\cite{bernardi2012developers}, social influence on contributions~\cite{tsay2014influence} and the project establishment~\cite{tsay2014influence}. First, we investigate how commits that introduce bugs (\textit{buggy} commits) differ from commits that are not involved in the introduction of bugs (\textit{clean} commits) in terms of these factors. Then, we analyze how strong is the difference between buggy and clean commits. Finally, we evaluate the effect of each factor on commit bugginess (\textit{i.e.}, the likelihood of a commit to introduce a bug) when considering the presence of multiple factors.

To perform our study, we collect data from eight open-source projects hosted on GitHub (see \textbf{Table~\ref{tab:projects})}. In particular, we collect 6,537 bug reports and compute 19 metrics. We use these metrics to characterize the factors analyzed. Our study led to four main findings: (i) both technical and social factors are able to discriminate between buggy and clean commits; (ii) technical factors are more determining to differentiate buggy commits from clean commits; (iii) the developers' habits of not following technical contribution norms and the developer's commit bugginess are associated with an increase on the likelihood of a new commit being buggy; and, finally, (iv) the project's establishment, the ownership level of developers' commit, and the social influence are related to a lower chance of introducing bugs. These findings shed light towards improving state-of-the-art techniques that may assist code reviewers during the inspection of bugs.


The remainder of this paper is structured as follows. Section~\ref{sec:study_design} presents the design of the empirical study, while its results are presented in Section~\ref{sec:results}. Section~\ref{sec:discussion} discuss our findings. Section~\ref{sec:threats} presents the threats to validity. Section~\ref{sec:related_work} discusses the related work on commit bugginess. Finally, Section~\ref{sec:conclusion} concludes this paper.

\section{Study Design}
\label{sec:study_design}

Open-source environments like GitHub enable developers with different technical capabilities and social interactions to contribute actively and simultaneously on the same software project. Also, developers may perform a variety of activities, for instance: push commits, open/close pull requests and issues, as well as discuss about contributions. Although, developers can collaborate on different projects, their technical capabilities and social interactions may be determining factors to the software quality. For example, a developer that has never communicated with others involved in a project may not have enough knowledge about it and, therefore, he may inadvertently introduce bugs when performing a commit. In this context, our study investigates the relation between technical, social factors and the introduction of bugs in open-source software projects. To do so, we define two research questions:

\smallskip
\textbf{RQ1:} \textit{Do bug-introducing commits differ from clean commits in terms of technical and social factors? How strong (or relevant) is the difference?}
\smallskip

Understanding which factors are more related to \textit{buggy} or clean commits may help code reviewers to avoid the introduction of bugs during the software development. For example, if the number of modified files in buggy commits is greater than clean commits, code reviewers may want to double check commits with a high number of modified files. Hence, \textbf{RQ1} aims at investigating if there is a statistically significant difference, and how strong the difference is~\cite{sullivan2012using}, between buggy and clean commits, by considering technical and social factors.

\smallskip
\textbf{RQ2:} \textit{In the presence of multiple factors, what factors most affect the bugginess of a commit?}
\smallskip

In the previous research question, we analyze which factors are able to discriminate between buggy and clean commits as well as how strong is this difference. However, during software development, the influence of different factors may contribute to commit bugginess~\cite{thongtanunam2016revisiting, tufano2017empirical, bernardi2012developers, posnett2013dual, eyolfson2011time}. These studies analyze the effects of factors in isolation, without considering the simultaneous influence of different factors. For example, the commit size and developer's experience may simultaneously contribute to commit bugginess. Hence, the \textbf{RQ2} aims at investigating the effect of each factor on commit bugginess by considering the presence of other factors.

\subsection{Technical and Social Factors}
\label{sec:factors}

To answer our research questions, we analyze five technical and three social factors. The former factors are related to technical contributions: developer's experience in a project (\textbf{F1}), developer's habit to follow well-known technical contribution norms (\textbf{F2}), nature of a developer's commits (\textbf{F3}), bugginess of a developer’s commits (\textbf{F4}), and project establishment (\textbf{F5}). The next three factors are social ones and capture interactions among developers in projects. We focus on ownership level of developers' commits (\textbf{F6}), comments performed by developers (\textbf{F7}) and their social influence (\textbf{F8}). We selected these factors since they were previously considered in studies involving open-source environments~\cite{tsay2014influence, gousios2014exploratory, dabbish2012social, eyolfson2011time, sliwerski2005changes, kamei2013large}. Each studied factor and their motivations are detailed below.

\subsubsection{Developers' Experience (F1)}
\label{sec:f1}

Although previous studies have assessed the influence of developer's experience on the likelihood of their commits being buggy~\cite{eyolfson2011time, mockus2010organizational, rahman2011ownership, tufano2017empirical}, they presented contradictory conclusions about this relation. While Eyolfson~\textit{et al.}~\cite{eyolfson2011time} and Rahman \& Devanbu~\cite{rahman2011ownership} show that experienced developers are less likely to introduce bugs, Mockus~\cite{mockus2010organizational} and Tufano~\textit{et al.}~\cite{tufano2017empirical} indicate that more experienced developers are more likely to introduce bugs. The contradictory findings indicated by these studies lead us to investigate again some of their analyses. However, these studies have assessed the influence of developers' experience by considering only their number of previous commits or the number of days in which they have been associated with a project.
Therefore, we also use data about the code review process to characterize the developers' experience. Hence, we analyze  \textbf{F1} to enhance the understanding on the relationship between developer's experience and commit bugginess.


\subsubsection{Developer's Habit to Follow Technical Contribution Norms (F2)}
\label{sec:f2}

Studies~\cite{dabbish2012social, tsay2014influence} indicate that project managers and code reviewers prefer to receive contributions (\textit{pull requests}) that follow certain norms, such as, the inclusion of tests, commits with a lower number of files changed and a higher legibility, their goal is to improve the software quality. Moreover, other studies~\cite{sliwerski2005changes, mockus2010organizational, kamei2013large} assessed quality measures associated with technical contribution norms (\textit{e.g.}, commits size and complexity) and their influence on commit bugginess. However, none of these studies assess the relation between the habit of following technical contribution norms and commit bugginess. For example, if the developer's commits include tests or involve only with small pieces of code; do these commits tend to insert fewer bugs? Hence, we analyze the \textbf{F2} factor, which wants to investigate the relation between following technical contribution norms and the commit bugginess.




\subsubsection{Nature of Developer's Commits (F3)}
\label{sec:f3}

Previous studies on commit bugginess~\cite{kamei2013large, guo2010characterizing, purushothaman2005toward} state that bug-fix commits are more likely to introduce a new bug in the software. This finding indicates that the nature of a commit may be a relevant indicator of bugginess. In addition, studies~\cite{hattori2008nature, mockus2000identifying, dragan2011using} provide commit classification strategies able to recognize the nature of a code change. Based on these strategies, we define the \textbf{F3} factor to evaluate the relation between the nature of developer's commits and their bugginess.


\subsubsection{Developer's Commits Bugginess (F4)}
\label{sec:f4}

Bugginess is the likelihood of a commit to introduce a bug. The bugginess of developer's commits is commonly used as a relevant factor in previous studies~\cite{eyolfson2011time, mockus2010organizational, rahman2011ownership, kamei2013large}. Eyolfson \textit{et al.}~\cite{eyolfson2011time} used the percentage of developer's commits that are buggy to determine how \textit{buggy} a developer's commits are. However, these studies did not evaluate if a developer's commit bugginess (\textit{i.e.}, the percentage of buggy commits) may influence the introduction of new bugs. For instance, is a developer whose commits are mostly \textit{buggy} more likely to introduce bugs in the future? Thus, we study \textbf{F4} to investigate the relation between developers' commits bugginess and the introduction of new bugs.


\subsubsection{Project Establishment (F5)}
\label{sec:f5}

Open-source projects are constantly evolving, growing and attracting new developers and followers, who, eventually, will demand an increase in the stability of the project~\cite{nakakoji2002evolution}. As a project becomes more stable, other important projects may become dependent on that project~\cite{dabbish2012social}. Thus, in light of these concerns, code reviewers of more established projects may be much more conservative and careful when accepting new contributions~\cite{tsay2014influence}. Hence, we define the \textbf{F5} factor, which aims at investigating whether the establishment of a project affects commit bugginess. 

\subsubsection{Ownership Level of Developers' Commits (F6)}
\label{sec:f6}

Prior work~\cite{rahman2011ownership, bird2011don, thongtanunam2016revisiting} have studied the ownership level of developers in a source code. Code ownership means how much a developer is responsible for the code committed to the repository. The lower the ownership level of a developer in a code, more developers are working on it and, consequently, the higher the interactions among those developers through the source code. These studies~\cite{rahman2011ownership, bird2011don, thongtanunam2016revisiting} focus only on the ownership of a particular source code entity, \textit{e.g.}, files. Our work has a commit granularity level, using the ownership of each file modified in that commit. There is no work analyzing the relationship between a developer's ownership in a commit level and commit bugginess. Therefore, in our study, we want to analyze if a developer that works mostly on his own code is less (or more) likely to introduce bugs.



\subsubsection{Communication with the Community of a Project (F7)}
\label{sec:f7}

A previous study~\cite{bernardi2012developers} analyzed the relation between the communication among developers and the introduction of bugs. The results suggest that bug-introducing committers discuss significantly less than other committers. In this study, the authors considered only the interaction among developers in the bug-tracking system (Bugzilla) of two projects. However, open-source environments~\cite{tsay2014influence, dabbish2012social} support the communication among developers in different ways, for instance, GitHub supports discussions about: (i) feature implementation and bug fixes through pull requests; (ii) report bugs or request features through issues; or (iii) changes made in a specific commit. Our intuition is that those different ways of communicating in a project community should be also considered when one investigates the effects on commit bugginess. Hence, \textbf{F7} considers all these communication forms as part of the conversation of a developer with the community. This way, we analyze the relation between commit bugginess and the amount of developer's communication with the community of a project hosted on GitHub.


\subsubsection{Social Influence on Contributions (F8)}
\label{sec:f8}

Along the evolution of a open-source project, its developers perform a variety of actions and contributions. These include: follow other developers, reviewing a source code, opening or closing pull requests or issues. Such actions and contributions are the mean in which developers play the roles (\textit{e.g}, core developers and reviewers). As a consequence, these developers become increasingly influential in the community.  Our goal is to investigate how this influence affects the commit bugginess of these developers.
    



\subsection{Metrics}
\label{sec:metrics}


\textbf{Developer's Experience (F1):} We use three metrics to characterize the developer's experience in a project. Our intuition is that \textit{the higher the value of these metrics, the more a developer understands the project and its source code, and, therefore, the more experienced he is}. To compute the values of each metric, we consider the interval between the first developer's commit to a project and the instants in which the commits authored by him were pushed. These metrics are detailed below:

\begin{itemize}[leftmargin=*]
    \item \textit{Experience (EXP)}: this metric quantifies experience as the number of commits authored by a developer~\cite{kamei2013large};
    
    \item \textit{Recent Experience (REXP)}: to measure the recent experience, we consider the developer's experience (\texttt{EXP}) weighted by the age of his previous changes, as defined by~\cite{kamei2013large}. By using the \texttt{REXP}, we give a higher weight to more recent changes. As a consequence, we can attribute more experience to developers who have contributed recently;

    \item \textit{File Experience (FEXP)}: it measures the developer's experience in the files modified in a commit authored by him. Particularly, for each file modified in a commit, we define the developer's experience as the number of previous commits authored by him on this file. Then, we compute this metric as the sum of the experience in each file.
    
\end{itemize}


\textbf{Technical Contribution Norms (F2):} We use five metrics to characterize the factor related to the technical contribution norms. Such metrics are based on the norms described in the work of Tsay at al.~\cite{tsay2014influence}, which analyzed the influence of the presence of tests (and other technical metrics) in commits on the acceptance of pull requests. In our study, we can analyze, for example, if \textit{the higher the presence of tests in the developer's commits, the more compliant with technical contribution norms the developer is}.

\begin{itemize}[leftmargin=*]
    \item \textit{Tests (\%) (TP)}: this metric measures the percentage of developer's commits that contain tests. Studies~\cite{tsay2014influence, dabbish2012social} indicate that reviewers prefer contributions containing tests because they are likely to be more reliable. Hence, we define the \textbf{TP} metric to measure how reliable developers' contributions are. To extract this metric, we adopt the procedure described in previous work~\cite{tsay2014influence} as the authors report a high accuracy. 
    
    \item \textit{Modified Files (MF)}: this metric counts the number of files modified in a commit. The motivation for this metric relies on studies~\cite{tsay2014influence, marlow2013impression} that indicate large contributions (involving a high number of modified files) are harder to understand or evaluate by project reviewers;
    
    \item \textit{Median of Modified Files (MMF)}: it measures the median of modified files among all the commits authored by a developer. This metric was defined to characterize the usual behavior (habit) of developers in terms of the number of files modified in their commits. By calculating the median of modified files, we can investigate, for instance, if developers that constantly modify many files are more (or less) likely to introduce bugs;
    
    \item \textit{Changed Lines (CL)}: it represents the number of changed lines in a commit. A changed line can consist of its addition or a deletion in a commit;
    
    \item \textit{Median of Changed Lines (MCL)}: it represents the median of changed lines among all the commits authored by a developer. 
\end{itemize}

We define the \texttt{CL} and \texttt{MCL} metrics as complementary metrics of \texttt{MF} and \texttt{MMF}, respectively, aiming at characterizing the legibility of developers' contributions.  We compute the \texttt{MF} and \texttt{CL} metrics by considering the instant that the commits were pushed in a project. To extract the \texttt{TP}, \texttt{MMF}, and \texttt{MCL} metrics, we adopted the same interval used to compute the experience metrics.

\textbf{Nature of Commits (F3):} We use four metrics to characterize the nature of developers' commits. These metrics are based on the classification of commit nature's described in Hattori \& Lanza~\cite{hattori2008nature}. The authors define four categories of commits based on a keyword analysis of the textual content in their messages: \textit{Forward Engineering}; \textit{Reengineering}; \textit{Corrective Engineering}; and \textit{Management}. Our intuition is that \textit{the higher the value of these metrics, the more a developer is focused on a specific commit nature}. The process of computing such metrics was the same adopted for the experience metrics. The measure of the metrics is the percentage of commits classified for each category, and named as: \textit{Forward Engineering (\%) (FEP)}; \textit{Reengineering (\%) (RP)}; \textit{Corrective Engineering (\%) (CEP)}; \textit{Management (\%) (MP)},

\textbf{Commits Bugginess (F4):} Inevitably, during the software development process, developers make changes that introduce bugs~\cite{sliwerski2005changes}. In this context, we analyze if the commits of developers who have previously introduced bugs are more likely to be buggy. To measure this factor, we evaluate the \textit{Percentage of Buggy Commits (PBC)} previously authored by a developer. Our intuition is that \textit{the higher the value of this metric, the more ``harmful'' developer's commits are}. The process of computing this metric is the same adopted for the developer's experience metrics.

\textbf{Project Establishment (F5):} We measure the establishment of a project as the \textit{Age} of a project since its first commit, \textit{i.e.}, how long (in days) a project has existed on GitHub. Such metric was defined by Tsay \textit{et al.}~\cite{tsay2014influence} and our intuition is that \textit{the higher the value of this metric, the more mature, and therefore, established, a project is}. We compute this metric at the instant that a commit was pushed.

\textbf{Ownership Level (F6):} We use two metrics to characterize the ownership level factor. Our intuition is that \textit{the higher the value of these metrics, the lower the interaction among developers in a source code.} We describe these metrics below:

\begin{itemize}[leftmargin=*]
    \item \textit{Commit Ownership (CO)}: this metric measures how much a developer ``owns'' the files modified in a commit that he authored. For each modified file, we measure the ownership of a developer in a file as $\frac{1}{d}$, where $d$ is the number of developers that previously authored a commit involving the file. Then, we define the ownership level of a developer in a commit as the median of the files ownerships. To compute this metric, we consider the instant that a commit was pushed to a project;
    
    \item \textit{Median Ownership (MO)}: it represents the median of the commits' ownerships among all the commits authored by a developer. The process of computing this metric is the same adopted for the developer's experience metrics.
\end{itemize}

\textbf{Communication within the Community (F7):} During the software development, developers can perform diverse activities on GitHub. For instance, they can communicate with the community by posting comments about different topics in issues and pull requests. Such interactions may represent the involvement of a developer on a project. In this context, we evaluate the \textit{Number of Comments (NC)} performed by a developer in a GitHub repository. Our intuition is that \textit{the higher the value of this metric, the more a developer is involved in a project}. The process of computing this metric is the same adopted when we compute the \textbf{F1} metrics.

\smallskip

\textbf{Social Influence on Contributions (F8):} We use five metrics to characterize the developer's social influence in a project. Our intuition is that \textit{the higher the value of these metrics, the higher the developer's influence in a project}.

\begin{itemize} [leftmargin=*]
    \item \textit{Code Review Experience (EXPRev)}: this metric indicates the developers' experience regarding their activities involving code review in GitHub projects. Particularly, we measure these activities in terms of the number of: (i) open and closed issues; (ii) open and closed pull requests; and (iii) comments on issues and pull requests;
    
    \item \textit{Recent Code Review Experience (REXPRev)}: this metric attributes more experience to developers whose performed code review activities recently. To do that, we analyze the code review experience (EXPRev) weighted by the age of previous review activities performed by the developer;

    \item \textit{Social Contributions (SC):} this metric represents the developer's contributions in terms of: (i) opened issues; (ii) issues that the developer opened and were closed; (iii) opened pull requests;
    (iv) pull requests the developer opened and were already closed; (v) pull requests that the developer opened and were merged;
    
    \item \textit{Social Distance (SD):} Defined by Tsay~\textit{et al.}~\cite{tsay2014influence}, this metric calculates the number of times whether or not the submitter (pull request owner) follows the developer that closed or merged his pull request;

    \item \textit{Median of Social Distance (MSD):} it measures the median of the social distance over time.  By calculating the median of the social distance, we can investigate if a higher `follow-relationship' in pull requests influences the bug introduction rate of a developer.
\end{itemize}

\subsection{Project Selection}
\label{sec:project_selection}

We manually select eight GitHub Java projects. The criteria to select the projects are: (i) the projects must be open-source and their changes history must be hosted on GitHub. This way, we ensure the full access to the software history; (ii) the projects must use the GitHub issues as the default bug-tracking tool. This way, we standardize our bug report analysis; (iii) the projects must be currently active and have been maintained or evolved for a long period of time. The main motivation to this criteria is to ensure that the projects are active and relevant to the GitHub community; and (iv) the projects must have a relevant number of reported bugs and involved developers. This way, we ensure that the projects have enough bug-related data to be investigated. Table~\ref{tab:projects} summarizes the characteristics of the projects, which have a high number of developers and a large number of commits and bugs associated with them. Moreover, all projects have thousands of GitHub \textit{stars}, that is a measure of community interest in a project~\cite{dabbish2012social}.


\begin{table}
    \caption{Selected GitHub Projects and their Characteristics}
    \label{tab:projects}
    \resizebox{\columnwidth}{!}{%
    \begin{tabular}{ccccccc}
        \toprule
        \textbf{Project} & \textbf{Domain} & \textbf{Commits} & \textbf{Bug Reports Collected} & \textbf{Developers} & \textbf{Stars} \\
        \midrule
        Elasticsearch    & Search Engine   & 45,747           & 2,305                & 1,197               & 40,540         \\
        Spring-boot      & App Builder     & 21,325           & 949                  & 567                 & 37,232         \\
        Netty            & Framework       & 9,263            & 1,134                & 369                 & 18,980         \\
        Bazel            & Build System    & 21,443           & 854                  & 437                 & 11,792         \\
        Presto           & Query Engine    & 15,489           & 296                  & 280                 & 8,972          \\
        RxJava           & Library         & 5,530            & 208                  & 241                 & 38,709         \\
        Signal-Android   & Messenger       & 4,134            & 509                  & 178                 & 11,382         \\
        OkHttp           & HTTP Client     & 3,869            & 282                  & 182                 & 32,023         \\
        \bottomrule
    \end{tabular}}
\end{table}

\subsection{Collecting Bug Reports}
\label{sec:bug_reports}

The GitHub issues are used to keep track of tasks, enhancements, and bugs related to a project. Furthermore, developers can associate labels with each issue to characterize it, such as ``bug'' or ``enhancement''. After fixing the bug or implementing an enhancement, the issue is closed. To collect the reports of fixed bugs in the selected projects, we mined the closed issues related to bugs (or defects) existing in each project. In order to identify these issues, we verified the ones containing the ``bug'' or ``defect'' labels, resulting in $6,537$ bug reports from the eight projects analyzed. Furthermore, we conducted a careful manual validation of the collected bug reports to guarantee that they are related to the report of bugs. After the manual validation process, we considered $5,034$ GitHub issues ($77\%$ of the total collected) that were classified as actual bug reports and investigated in our study.

\subsection{Locating Bug-introducing Changes}
\label{sec:locating_bugs}

During software development, developers make changes in the source code that may introduce bugs. We will further refer to these changes as \textit{bug-introducing changes}~\cite{kim2006automatic}. To locate these, we implement the \textit{SZZ} algorithm~\cite{sliwerski2005changes}, which aims at identifying the commits that introduced a bug in a software project. In this context, the SZZ algorithm requires the commit(s) 
that fixed a given bug, hence, we use a GitHub strategy to identify these commits. Keywords regarding fixing and closing issues followed by an issue number will automatically close the issue. In this way, we assume the commit that closed the issue as being the commit that fixed the bug.

We employ the SZZ algorithm for each validated bug report from the eight selected projects. As a result, we obtain a total of $10,674$ unique candidate bug-introducing changes. In addition, we conducted a careful manual validation on a sample of $922$ bug-introducing changes reported by SZZ. This sample presents a margin of error of $4\%$ at a confidence level of $99\%$. Four pairs of researchers (familiar with the SZZ algorithm) from our research lab validated this sample. Each pair was responsible for a fraction of the sample, and each individual validated the same candidate bug-introducing changes. To take into account possible subjectivity during the validation, the researchers classified each candidate into four levels of certainty: ``low'', ``moderate'', ``high'', and ``absolute''. In the cases the researchers of a pair classified a candidate very differently (\textit{e.g.}, ``low'' and ``absolute''), one of the authors solved   the conflict. After the validation procedure, we obtained an accuracy of $65\%$ for our SZZ implementation. We conducted the validation due to the high numbers of false-positives (changes reported as bug-introducing when they are actually not) reported in previous studies~\cite{kim2006automatic, williams2008szz, da2017framework}.


\subsection{Data Collection}
\label{sec:data_colletction}

To collect the data used to compute the metrics related to technical and social factors (Section~\ref{sec:metrics}), we use the GitHub API. First, we collect the public identifier (\textit{username}) on GitHub of the developers that authored at least one commit in each of the studied projects. Then, we mine all the commits, issues, pull requests, and comments of a project. Finally, we link the collected authors to the activities (\textit{e.g.}, commits, pull requests) that they performed on the project, through their \textit{usernames}. As a result of this process, we obtain data about $2,949$ developers, which authored at least one commit in the repository. Moreover, we collect $96,258$ commits, $33,244$ pull requests, $45,554$ issues, and $358,200$ comments related to the eight projects analyzed.

\subsection{Data Analysis}
\label{sec:data_analysis}

To answer \textbf{RQ1}, we use the \textit{Wilcoxon Rank Sum Test}~\cite{whitley2002statistics} to verify which metrics are able to discriminate between buggy and clean commits. This non-parametric (\textit{i.e.}, normality is not required~\cite{mcdonald2009handbook}) test allows us to decide whether two populations (in our study, metrics related to the buggy and clean commits of each project) are \textit{identical} or not. To ensure statistical significance, we adopt the customary $.05$ significance level (\textit{p-value} $< 0.05$) for this test. Since we are performing multiple comparisons, we need to adjust the \textit{p-values} of the \textit{Wilcoxon Tests} to take into account the increased chance of rejecting the \textit{null hypothesis} simply due to random chance. To do so, we apply the widely used \textit{Bonferroni correction}~\cite{mcdonald2009handbook}, which controls the \textit{familywise} error rate. For this method, we consider that each project is a family, which means that we perform the correction in the \textit{p-values} of the metrics at the project level.

Furthermore, in \textbf{RQ1} we also use the \textit{Cliff's Delta (d)} measure~\cite{grissom2005effect} to evaluate how strong is the difference between buggy and clean commits in terms of the metrics analyzed. Similarly to the \textit{Wilcoxon Rank Sum} test, the \textit{Cliff's Delta (d)} is a non-parametric effect size measure. In order to interpret the Cliff's Delta (\textit{d}) effect size, we employ a well-known classification~\cite{romano2006exploring}. It defines four categories of magnitude: \textit{negligible}, \textit{small}, \textit{medium}, and \textit{large}.

To answer the \textbf{RQ2}, we evaluate the effect of each factor in the presence of the other ones. To do so, we create a \textit{multiple logistic regression} model for each studied project, where each metric is a predictor and the outcome variable is whether a commit is \textit{buggy} or not. In other words, we create a regression model that predicts the likelihood of commit bugginess. We choose a \textit{multiple logistic regression} approach due to the fact that we are studying the effect of multiple predictors (\textit{i.e.}, the metrics) in a binary response variable.

We report the effect of the metrics in the likelihood of a commit being buggy in terms of \textit{odds ratios}. \textit{Odds ratios} are the increase or decrease in the odds of a commit being buggy occurring per ``unit'' value of a predictor (metric). An odds ratio $<1$ indicates a decrease in these odds, while $>1$ indicates an increase. Since our metrics are heavily skewed (\textit{e.g.}, only a few developers perform most of the commits in a project~\cite{steinmacher2018almost}), we apply a $\log_{2}$ transformation on the right-skewed predictors and a $x^{3}$ transformation on the left skewed.

To ensure that multicollinearity would not affect our model, we remove the metrics which have a pair-wise correlation coefficient above $0.7$~\cite{dormann2013collinearity}. Moreover, to ensure normality, we normalize the continuous predictors in the model. As a result, the mean of each predictor is equaled to zero and the standard deviation to one. Finally, to ensure statistical significance of the predictors, we employ the customary $.05$ significance level for each predictor in the regression model.

We present the results of our analysis based on the 10,674 unique candidate bug-introducing changes reported by the SZZ algorithm. However, we also perform the same analysis on the sample validated by the four pairs of researchers, containing the $922$ bug-introducing changes. The results of both analysis are similar, as can be seen in our replication package.


\subsection{Replication Package}
\label{sec:replication}

We provide a replication package,\footnote{\url{https://filipefalcaos.github.io/saner-2020}} that contains the source code used in this study, the data used in the analyses, a detailed example of the methodology, a deeper characterization 
of our dataset, and details on the validation regarding the SZZ algorithm (see Section~\ref{sec:locating_bugs}).

\section{Results}
\label{sec:results}

The following section present the results for the two research questions (Section~\ref{sec:study_design}).

\smallskip
\textbf{RQ1:} \textit{Do bug-introducing commits differ from clean commits in terms of technical and social factors? How strong (or relevant) is the difference?}
\smallskip

Table~\ref{tab:wilcoxon_delta} presents the results of \textbf{RQ1}. 
The first and seconds columns represent the technical and social factors, and the metrics associated with them. The remaining columns describe the metrics and their respective magnitudes of the Cliff's Delta related to each studied project. We use the $(+)$ symbol to indicate if \textit{d} was positive, and $(-)$ otherwise. In addition, we use four levels to measure the strength of a magnitude: (small) *; (medium) **; (large) ***; and (negligible). We do not use a symbol to represent this last level.  The cells in gray represent the \textit{p-values} of the metrics that obtained statistical significance in the \textit{Wilcoxon Rank Sum} test. Each factor is assessed in terms of N cases where N is given by the number of metrics multiplied by 8 (number of projects).

\begin{table}
    \centering
    \caption{Statistical Significance (\textit{p-value}) of the \textit{Wilcoxon Rank Sum Test} and the Cliff's Delta (\textit{d}) Magnitude Classification}
    \label{tab:wilcoxon_delta}
    \resizebox{\linewidth}{!}{%
    \setlength{\tabcolsep}{3pt}
    \begin{tabular}{>{\bfseries}ccllllllll}
        \toprule
        Factor              & \textbf{Metric} & \textbf{Elastic} & \textbf{Spring} & \textbf{Netty} & \textbf{Bazel} & \textbf{Presto} & \textbf{RxJava} & \textbf{Signal} & \textbf{OkHttp} \\
        \midrule
        \multirow{3}{*}{F1} & EXP             & \gc{$(+)$}       & \gc{$(-)$ **}   & \gc{$(+)$ *}   & \gc{$(-)$}     & $(-)$           & \gc{$(+)$ *}    & \gc{$(-)$ *}    & $(-)$           \\
                            & FEXP            & $(+)$            & \gc{$(-)$}      & \gc{$(+)$}     & $(+)$          & \gc{$(+)$}      & $(+)$           & $(-)$           & $(+)$           \\
                            & REXP            & \gc{$(-)$}       & \gc{$(-)$ **}   & \gc{$(+)$ *}   & \gc{$(-)$ *}   & \gc{$(-)$}      & \gc{$(+)$ *}    & \gc{$(-)$ *}    & $(-)$           \\
                            
        \midrule
        \multirow{5}{*}{F2} & TP              & $(+)$            & $(-)$           & \gc{$(+)$}     & $(-)$          & $(+)$           & \gc{$(+)$ *}    & \gc{$(-)$ *}    & \gc{$(+)$}      \\
                            & MCL             & \gc{$(+)$ *}     & \gc{$(+)$ **}   & \gc{$(+)$}     & $(+)$          & \gc{$(+)$ *}    & \gc{$(+)$ **}   & \gc{$(+)$ **}   & \gc{$(+)$ *}    \\
                            & CL              & \gc{$(+)$ ***}   & \gc{$(+)$ ***}  & \gc{$(+)$ ***} & \gc{$(+)$ ***} & \gc{$(+)$ ***}  & \gc{$(+)$ ***}  & \gc{$(+)$ ***}  & \gc{$(+)$ ***}  \\
                            & MMF             & \gc{$(+)$}       & \gc{$(+)$ **}   & $(-)$          & $(+)$          & \gc{$(+)$ *}    & \gc{$(+)$ **}   & \gc{$(+)$ *}    & \gc{$(+)$ *}    \\
                            & MF              & \gc{$(+)$ **}    & \gc{$(+)$ ***}  & \gc{$(+)$ **}  & \gc{$(+)$ **}  & \gc{$(+)$ ***}  & \gc{$(+)$ ***}  & \gc{$(+)$ ***}  & \gc{$(+)$ **}   \\
        
        \midrule
        \multirow{4}{*}{F3} & RP              & \gc{$(-)$}       & \gc{$(+)$ *}    & $(+)$          & \gc{$(-)$ *}   & $(-)$           & \gc{$(-)$ *}    & $(-)$           & \gc{$(+)$ *}    \\
                            & CEP             & \gc{$(+)$}       & \gc{$(+)$ *}    & \gc{$(+)$}     & \gc{$(-)$ *}   & $(-)$           & \gc{$(+)$ **}   & $(-)$           & $(+)$           \\
                            & FEP             & \gc{$(+)$}       & \gc{$(+)$}      & \gc{$(-)$}     & \gc{$(-)$}     & $(+)$           & \gc{$(-)$ **}   & \gc{$(+)$}      & \gc{$(+)$ *}    \\
                            & MP              & \gc{$(-)$}       & \gc{$(-)$ **}   & \gc{$(-)$}     & \gc{$(+)$ *}   & $(+)$           & $(-)$           & $(+)$           & \gc{$(-)$ *}    \\
        
        \midrule
        \multirow{1}{*}{F4} & PBC             & \gc{$(+)$ **}    & \gc{$(+)$ **}   & \gc{$(+)$ *}   & \gc{$(+)$ **}  & \gc{$(+)$ *}    & \gc{$(+)$ ***}  & \gc{$(+)$ **}   & \gc{$(+)$ *}    \\
        
        \midrule
        \multirow{1}{*}{F5} & Age             & \gc{$(-)$ *}     & \gc{$(-)$ ***}  & \gc{$(+)$}     & \gc{$(-)$ **}  & \gc{$(-)$ *}    & \gc{$(+)$ **}   & \gc{$(-)$ **}   & \gc{$(-)$}      \\
                            
        \midrule
        \multirow{2}{*}{F6} & MO              & \gc{$(+)$ *}     & \gc{$(+)$ *}    & \gc{$(+)$}     & \gc{$(+)$ *}   & $(-)$           & $(-)$           & \gc{$(+)$}      & $(-)$           \\
                            & CO              & \gc{$(+)$}       & \gc{$(+)$ *}    & \gc{$(-)$}     & \gc{$(+)$}     & $(-)$           & \gc{$(+)$ **}   & $(-)$           & \gc{$(+)$ *}    \\

        \midrule
        \multirow{1}{*}{F7} & NC              & \gc{$(-)$}       & \gc{$(-)$ **}   & \gc{$(+)$ *}   & $(+)$          & \gc{$(-)$}      & \gc{$(+)$ **}   & \gc{$(-)$ *}    & $(-)$           \\
        
        \midrule
        \multirow{5}{*}{F8} & EXPRev          & \gc{$(-)$}       & \gc{$(-)$ **}   & \gc{$(+)$ *}   & $(+)$          & \gc{$(-)$}      & \gc{$(+)$ **}   & \gc{$(-)$ *}    & $(-)$           \\
                            & REXPRev         & \gc{$(-)$}       & \gc{$(-)$ **}   & \gc{$(+)$ *}   & \gc{$(-)$ *}   & \gc{$(-)$}      & \gc{$(+)$ *}    & \gc{$(-)$ *}    & $(-)$           \\
                            & SC              & \gc{$(-)$}       & \gc{$(-)$ **}   & \gc{$(+)$ *}   & $(-)$          & $(-)$           & \gc{$(+)$ **}   & \gc{$(-)$ *}    & $(-)$           \\
                            & MSD             & \gc{$(-)$ *}     & \gc{$(-)$ **}   & \gc{$(-)$}     & $(+)$          & $(-)$           & \gc{$(+)$ ***}  & \gc{$(-)$ *}    & $(-)$           \\
                            & SD              & \gc{$(-)$}       & \gc{$(-)$ **}   & \gc{$(+)$}     & $(+)$          & $(-)$           & \gc{$(+)$ ***}  & \gc{$(-)$ *}    & $(-)$           \\
        \bottomrule
    \end{tabular}}
\end{table}

We observe that the developers' experience metrics (\textbf{F1}) obtain a statistically significant difference in 16 out of 24 cases analyzed ($66\%$). The \texttt{FEXP} metric presents statistical significance only in three projects. The \texttt{EXP} metric does not obtain statistical significance only in two projects. The \texttt{REXP} metric presents even better results, not obtaining a statistical significance only in one project. Regarding the technical contribution norms metrics (\textbf{F2}), they obtain statistical significance in a number of cases greater than the \textbf{F1} metrics. The \textbf{F2} metrics obtain a statistically significant difference in $82\%$ of the cases analyzed. The \texttt{CL} and \texttt{MF} metrics present statistical significance in all projects, while \texttt{MCL} is not statistically significant only in \textit{Bazel}.

Regarding the \textbf{F3} factor, its metrics present statistical significance in 22 out of 32 cases analyzed ($68\%$). Particularly, the \texttt{FEP} metric obtains statistical significance in seven out of eight projects analyzed. Similarly to \texttt{FEP} (\textbf{F3}), the \texttt{REXP} (\textbf{F1}), \texttt{MCL} (\textbf{F2}), \texttt{CL} (\textbf{F2}), \texttt{MF} (\textbf{F2}), and \texttt{PBC} (\textbf{F4}) metrics also obtain statistical significance in at least seven of the projects analyzed. The \texttt{Age} (\textbf{F5}) metric reaches even better results by obtaining statistical significance in all projects analyzed. \textbf{Such results indicate that technical factors can discriminate between buggy and clean commits}.

When observing the social factors (\textbf{F6-F8}), both metrics characterizing the \textbf{F6} factor present a statistical significance in the majority of the projects. While the \texttt{CO} metric obtains a statistically significant difference in six out of eight projects, the \texttt{MO} metric is not statistically significant in three.  Also, we observe that the \texttt{NC} (\textbf{F7}) metric presents a statistical significance in six projects. Regarding the \textbf{F8} factor, the metrics present statistical significance in five out of the eight projects. On the other hand, no metric from this factor could reach statistical significance in the \textit{OkHttp} project. Note also that only \texttt{EXPRev} and \texttt{REXPRev} are statistically significant in \textit{Presto} and only one metric presents statistical significance in \textit{Bazel}. Similarly to the technical factors, \textbf{the social factors can also discriminate between buggy and clean commits}.

When we consider both the technical and social factors, they obtain a statistically significant difference in $75\%$ of the cases investigated. While the technical metrics present statistical significance in 78\% of the cases, the social metrics present in 70\%. Differently from technical metrics, no social metric could reach a statistically significant difference in all the projects analyzed. Note that the technical metrics  \texttt{CL}, \texttt{MF}, \texttt{PBC}, and \texttt{Age} present statistical significance in all projects.   

\smallskip
\begin{opus-box}
     Both technical and social factors can be used as a good guideline to distinguish between clear and buggy commits. However, the technical factors present a proportion of statistical significance cases greater than the social ones (78\% and 70\%, respectively).
\end{opus-box}

We previously analyzed if there is a statistically significant difference between clean and buggy commits in terms of technical and social factors. From now on, we also investigate the magnitude of this difference. 

\textbf{Positive Magnitude.} Notice that the technical metrics \texttt{MCL}, \texttt{MF}, \texttt{CL} (\textbf{F2}), and \texttt{PBC} (\textbf{F4}) present a positive magnitude in all projects analyzed. Indeed, we observe that the \texttt{CL} (Changed Lines) metric obtains a \textit{large} magnitude in all projects analyzed. The \texttt{MF} (Modified Files) and \texttt{PBC} (Percentage of Buggy Commits) metrics obtain a \textit{strength} slightly lower than \texttt{CL} by reaching magnitudes between \textit{small} and \textit{large}. We also observe that both the \texttt{MCL} (Median of Changed Lines) and \texttt{MMF} (Median of Modified Files) reach magnitudes that varied from \textit{negligible} up to \textit{medium}. While \texttt{MCL} presents a positive magnitude in all projects, the \texttt{MMF} presents in seven ones. Finally, all the \textbf{F8} metrics present a small (once), medium (twice), or large (twice) positive magnitudes in \textit{RxJava}. 

\textbf{Negative Magnitude.} Regarding the \textit{negative} magnitudes, the \texttt{Age} metric presents negative magnitudes in six projects, reaching a large one in \textit{Spring-boot}. Similarly to the \texttt{Age} metric, the \texttt{REXP} (\textbf{F1}) and \texttt{REXPRev} (\textbf{F8}) metrics also present negative magnitudes in six projects. However, they obtain a \textit{strength} equal or slightly lower than \texttt{Age} in such projects. We also observe that the \textbf{F1} and \textbf{F8} metrics present a negative magnitude varying from negligible to medium in the \textit{Spring-boot} and \textit{Signal-Android}, respectively. 

By analyzing both technical and social factors, the metrics present positive magnitude in 54\% of the cases. \textbf{If we consider only the statistically significant cases, the metrics present positive magnitude in 60\% of the cases}. Also, in the statistically significant cases, \textbf{the technical metrics CL, MF and PBC present a small or large positive magnitude in all the projects analyzed.} This means that the buggy commits of these projects have a significant higher CL, MF and PBC than clean commits. \textbf{Differently from the technical factors, we do not observe a tendency  of positive/negative magnitude in the social metrics.}

\smallskip
\begin{opus-box}
    There are strong differences between clean and buggy commits in terms of technical and social factors.  Moreover, the positive magnitudes are more frequent than the negative ones. Again, the technical factors are more determining than the social ones.
\end{opus-box}

\smallskip
\textbf{RQ2:} \textit{In the presence of multiple factors, what factors most affect the bugginess of a commit?}
\smallskip

In the \textbf{RQ1}, we analyze the metrics individually. Now, we use the \textit{odds ratios} technique to investigate the effect of each metric on commit bugginess in the presence of other metrics analyzed in our study. Table~\ref{tab:model} summarizes the effects of the metrics on commit bugginess in each project. The first column represents the technical and social factors. The remaining columns describe the metrics and their respective odds ratios related to the projects. For each project, we consider only the metrics that do not have collinearity among them. Like in \textbf{RQ1}, cells in gray represent the statistical significance. In addition, we use the $\uparrow$ symbol to indicate a risk-increasing effect, and the $\downarrow$ symbol otherwise. We analyze the risk-increasing or decreasing effect of each metric by considering only the statistically significant cases (cells in gray).

\begin{table*}
    \centering
    \caption{Multiple Logistic Regression Model of Commit Bugginess}
    \label{tab:model}
    \resizebox{\linewidth}{!}{%
    \setlength{\tabcolsep}{3pt}
    \begin{tabular}{>{\bfseries}cllllllll}
        \toprule
         Factor    & \textbf{Elasticsearch}        & \textbf{Spring-boot}          & \textbf{Netty}               & \textbf{Bazel}                   & \textbf{Presto}              & \textbf{RxJava}             & \textbf{Signal-Android}      & \textbf{OkHttp}                  \\
        \midrule                                                                                                                                                                                                                                                                                                     
        \multirow{2}{*}                    
        {F1}       & \gc{FEXP (0.96)} $\downarrow$ & \gc{FEXP (0.86)} $\downarrow$ & FEXP (1.01) $\uparrow$       & FEXP (1.04) $\uparrow$           & FEXP (1.04) $\uparrow$       & \gc{FEXP (1.20)} $\uparrow$ & FEXP (1.05) $\uparrow$       & FEXP (1.03) $\uparrow$           \\
      		       & EXP (0.93) $\downarrow$       &                               & EXP (1.06) $\uparrow$        &                                  &                              & EXP (1.44) $\uparrow$       &                              &                                  \\
        \midrule                                                                                                                                                                                                                                                                                                      
        \multirow{5}{*}             
        {F2}       & \gc{TP (1.36)} $\uparrow$     & \gc{TP (1.14)} $\uparrow$     & \gc{TP (1.24)} $\uparrow$    & TP (1.08) $\uparrow$             & \gc{TP (0.85)} $\downarrow$  & TP (0.83) $\downarrow$      &                              & \gc{TP (1.36)} $\uparrow$        \\
				   & MCL (0.98) $\downarrow$       & \gc{MCL (1.19)} $\uparrow$    &                              & MCL (1.08) $\uparrow$            & MCL (1.10)  $\uparrow$       & MCL (1.12) $\uparrow$       & MCL (1.15) $\uparrow$        & MCL (1.01) $\uparrow$            \\
                   & \gc{CL (2.83)} $\uparrow$     & \gc{CL (6.69)} $\uparrow$     & \gc{CL (3.36)} $\uparrow$    & \gc{CL (2.48)} $\uparrow$        & \gc{CL (2.94)} $\uparrow$    & \gc{CL (1.67)} $\uparrow$   & \gc{CL (2.97)} $\uparrow$    & \gc{CL (2.47)} $\uparrow$        \\
                   & MMF (0.95) $\downarrow$       & \gc{MMF (0.83)} $\downarrow$  & \gc{MMF (0.85)} $\downarrow$ & \gc{MMF (0.85)} $\downarrow$     &                              & MMF (1.07) $\uparrow$       & MMF (0.96)  $\downarrow$     &                                  \\
                   & MF (1.02) $\uparrow$          & MF (1.04) $\downarrow$        & MF (0.99) $\downarrow$       & \gc{MF (1.32)} $\uparrow$        & \gc{MF (1.16)} $\uparrow$    & \gc{MF (1.86)} $\uparrow$   & MF (1.20) $\uparrow$         & \gc{MF (1.23)} $\uparrow$        \\
        \midrule
        
        \multirow{4}{*}                                        
        {F3}       & \gc{RP (0.91)} $\downarrow$   & RP (1.03) $\uparrow$          & \gc{RP (0.87)} $\downarrow$  & RP (1.13) $\uparrow$             & RP (1.02)  $\uparrow$        & RP (1.28) $\uparrow$        & RP (0.88) $\downarrow$       & RP (1.09) $\uparrow$             \\
                   & \gc{CEP (1.07)} $\uparrow$    & CEP (0.91) $\downarrow$       & CEP (0.93) $\downarrow$      & \gc{CEP (0.79)} $\downarrow$     & CEP (1.03) $\uparrow$        & CEP (1.01) $\uparrow$       & CEP (0.92) $\downarrow$      & CEP (1.11) $\uparrow$            \\
                   & \gc{FEP (1.14)} $\uparrow$    & \gc{FEP (1.13)} $\uparrow$    & \gc{FEP (1.20)} $\uparrow$   & FEP (1.05) $\uparrow$            & \gc{FEP (0.87)} $\downarrow$ & FEP (1.03) $\uparrow$       & FEP (0.99) $\downarrow$      & FEP (1.05) $\uparrow$            \\
                   & \gc{MP (0.91)} $\downarrow$   & MP (0.92) $\downarrow$        & \gc{MP (0.87)} $\downarrow$  &                                  & MP (1.02) $\uparrow$         & MP (0.93) $\downarrow$      & MP (0.96) $\downarrow$       & \gc{MP (0.66)} $\downarrow$      \\
        \midrule
        
        \multirow{1}{*}
		{F4}       & \gc{PBC (1.39)} $\uparrow$    & PBC (0.99) $\downarrow$       & \gc{PBC (1.38)} $\uparrow$   & \gc{PBC (1.50)} $\uparrow$       & \gc{PBC (1.60)} $\uparrow$   &                             & PBC (1.07) $\uparrow$        & PBC (0.80) $\downarrow$          \\
        \midrule
        
        \multirow{1}{*}
		{F5}       &                               & \gc{Age (0.24)} $\downarrow$  & \gc{Age (1.53)} $\uparrow$   & \gc{Age (0.18)} $\downarrow$     & \gc{Age (0.41)} $\downarrow$ & \gc{Age (1.71)} $\uparrow$  & \gc{Age (0.25)} $\downarrow$ & Age (1.12) $\uparrow$            \\
		\midrule
        
        \multirow{2}{*}
        {F6}       & \gc{CO (1.06)} $\uparrow$     & CO (0.92) $\downarrow$        & \gc{CO (0.74)} $\downarrow$  & \gc{CO (0.87)} $\downarrow$      & \gc{CO (0.84)} $\downarrow$  & \gc{CO (2.13)} $\uparrow$   & \gc{CO (0.75)} $\downarrow$  & CO (1.11) $\uparrow$             \\
                   & \gc{MO (1.14)} $\uparrow$     & MO (1.02) $\uparrow$          &                              & MO (1.04) $\uparrow$             & \gc{MO (0.64)} $\downarrow$  & \gc{MO (0.34)} $\downarrow$ &                              &                                  \\
        \midrule
        
        \multirow{1}{*}
		{F7}       & NC (0.99) $\downarrow$        &                               &                              & NC (1.02) $\uparrow$             & \gc{NC (1.49)} $\uparrow$    &                             &                              &                                  \\
        \midrule
        
        \multirow{2}{*}
        {F8}       & \gc{MSD (0.76)} $\downarrow$  & SD (0.95) $\downarrow$        & \gc{MSD (0.74)} $\downarrow$ & \gc{MSD (1.11)} $\uparrow$       &                              &                             & MSD (1.17) $\uparrow$        &                                  \\
				   &                               &                               &                              & \gc{REXPRev (0.74)} $\downarrow$ &                              &                             &                              & \gc{REXPRev (0.68)} $\downarrow$ \\
        \bottomrule
    \end{tabular}}
\end{table*}

\textbf{Risk-increasing Effect.} We observe that the metrics Test Presence \textbf{(\texttt{TP})}, Changed Lines \textbf{(\texttt{CL})}, and Modified Files \textbf{(\texttt{MF})}, from \textbf{(F2)}; Forward Engineering \textbf{(\texttt{FEP})}, from \textbf{(F3)}; and Percentage of Buggy Commits \textbf{(\texttt{PBC})}, from \textbf{(F4)}; present a risk-increasing tendency. Note that only the \textbf{F2} factor obtains at least one metric (\texttt{CL}) having a effect in all projects analyzed. In particular, the \texttt{CL} metric obtains a risk-increasing effect in all cases. Among all the metrics analyzed in our study, \texttt{CL} reaches the highest effect in seven of the eight projects, reaching an odds ratio up to $6.69$ in \textit{Spring-boot}. Such fact indicates that each unit of the \texttt{CL} metric increases the odds of a commit being buggy by a factor of $1.67$ (\textit{RxJava}) up to $6.69$ (\textit{Spring-boot}). Similarly, the metrics \texttt{PBC}, \texttt{MF}, and \texttt{TP} also present a risk-increasing effect, but only in four projects. This high number of risk-increasing effects obtained by \texttt{TP} introduces some questions concerning code reviewers that prefer contributions containing tests~\cite{tsay2014influence}. In the next section, we discuss some issues that may lead the \texttt{TP} to obtain this tendency. Regarding the \texttt{FEP} metric, it presents a risk-increasing effect in three out of the four projects. Finally, two metrics, \texttt{MCL} (\textbf{F2}) and \texttt{NC} (\textbf{F7}), present only one statistically significant risk-increasing effect, which is not enough to characterize a tendency.



\textbf{Risk-decreasing Effect.} Differently from the previous metrics, the Median of Modified Files \textbf{(\texttt{MMF})}, from Technical Contribution Norms \textbf{(F2)}; Management \textbf{(\texttt{MP})}, from Nature of Commits \textbf{(F3)}; \textbf{\texttt{Age}}, from Project Establishment \textbf{(F5)}; and Commit Ownership \textbf{(\texttt{CO})}, from Ownership Level \textbf{(F6)}; present a risk-decreasing tendency. The \texttt{Age} metric reaches the highest risk-decreasing effect in four out of the six projects in which it is significant. It presents an odds ratio of $0.18$ in the \textit{Bazel} project, which means that each unit of \texttt{Age} decreases the odds of a commit being buggy by a factor of $0.18$ in this project. Similarly to the \texttt{Age} metric, the \texttt{CO} metric also obtains significant risk-decreasing effect in four out of six projects. Although this metric presents a risk-decreasing tendency, it reaches the highest risk-increasing effect in \textit{RxJava}. We discuss in more detail such case in Section~\ref{sec:discussion}. By analyzing the \texttt{MMF} and \texttt{MP} metrics, we note that they present a risk-decreasing  effect in all the three projects where each obtain statistical significance. 

The File Experience (\texttt{FEXP}), Median of Social Distance (\texttt{MSD}), and Median of Ownership (\texttt{MO}) metrics present a risk-decreasing effect in two out of the three projects they reached statistical significance. However, each of these also present one statistically significant risk-increasing effect. Finally, the \texttt{RP} metric obtain a statistical significance only in two projects, presenting a risk-decreasing effect. Due to the small amount of significant cases, the \texttt{FEXP}, \texttt{MSD}, \texttt{MO}, and (\texttt{RP}) metrics do not present a tendency towards a risk-decreasing effect.

When we analyze the risk of the metrics in terms of the technical and social factors, we observe that most of the metrics that compose the \textbf{F2} factor, as well as \textbf{F4}, present a tendency towards a risk-increasing effect. On the other hand, most of the metrics that compose the \textbf{F5}, \textbf{F6}, and \textbf{F8} factors show a risk-decreasing tendency.
 

\smallskip
\begin{opus-box}
    The Developer's Habit (F2) and Commit Bugginess (F4) factors present a risk-increasing tendency. On the other hand, Project Establishment (F5), Ownership Level (F6), and Social Influence on Contributions (F8) present a risk-decreasing tendency.
\end{opus-box}

\section{Discussion}
\label{sec:discussion}


\textbf{The contradictory effects of experience metrics.} Regarding the developers' experience, we observe that two metrics (\texttt{EXP} and \texttt{REXP}) reach a \textit{medium} and \textit{small} negative magnitude in the \textit{Spring-boot} and \textit{Signal-Android} projects, respectively (see Table~\ref{tab:wilcoxon_delta}). On the other hand, the same metrics reach a \textit{small} positive magnitude in the \textit{Netty} and \textit{RxJava}. Such contradiction reinforces the results discussed by previous studies~\cite{eyolfson2011time, mockus2010organizational, tufano2017empirical, rahman2011ownership}. Eyolfson \textit{et al.}~\cite{eyolfson2011time} provide evidence from two open-source projects that more experienced developers are less likely to introduce bugs. On the other hand, prior work~\cite{mockus2010organizational, tufano2017empirical} explains that more experienced developers are more likely to introduce bugs due to the complexity of their tasks. Thus, to understand why the effects of the experience factor is contradictory, we further investigate the relation of this factor with the complexity of the changes performed by the developers. In particular, we investigate whether more experienced developers perform more complex changes. To do so, we use the \textit{Spearman $\rho$ rank}~\cite{mcdonald2009handbook} technique to evaluate if there is a correlation between the developers' experience metrics and the complexity of their commits (\textit{i.e.}, median of changed lines in the previous commits).

We perform our investigation in the \textit{Spring-boot} and \textit{RxJava} projects since two experience metrics (\texttt{EXP} and \texttt{REXP}) present opposite effects in these projects. We use the classification defined by Cohen~\cite{cohen2013statistical} to determine the strength of the correlations between these two metrics and the commits complexity. In the \textit{Spring-boot}, the two experience metrics present only \textit{negative} correlations ranging from $-0.21$ (\textit{small}) up to $-0.53$ (\textit{large}). Such result suggests that more experienced developers usually perform less complex changes in the \textit{Spring-boot}. Such finding may be an indication why more experienced developers introduce fewer bugs in this project. On the other hand, the same two experience metrics presented positive correlations ranging from $0.27$ (\textit{small}) up to $0.45$ (\textit{medium}). Such result indicates that more experienced developers usually perform more complex changes in the \textit{RxJava}. Such finding may be an indication why more experienced developers introduce more bugs in this project. Even though these results shed light in the understanding about why experience metrics presented contradictory effects on commit bugginess, it is still necessary further analysis in order to obtain relevant conclusions, that will be investigated in future works.

\textbf{The risk-increasing effect of the \texttt{TP} metric.} Previous work~\cite{tsay2014influence, dabbish2012social} show that code reviewers prefer contributions that contain tests aiming at improving the software quality. However, results from \textbf{RQ2} (see Table~\ref{tab:model}) show that the percentage of developer's commits that contain tests (\texttt{TP}) has a risk-increasing tendency on commit bugginess. This finding sounds contradictory to the assumption that testing practices would improve the software quality. Thus, to better understand the reason behind this tendency, we further investigate the relation of the \texttt{TP} metric with the complexity of the changes performed by developers. Our intuition is that this metric may have a risk-increasing effect if the developers who constantly write tests in their commits also perform more complex changes. Similarly to the previous discussion, we use the \textit{Spearman $\rho$ rank} correlation to evaluate the relation between the \texttt{TP} metric and the complexity of their commits (\textit{i.e.}, median of changed lines in the previous commits).

We perform our investigation by using the \textit{Netty}, \textit{Spring-boot}, \textit{OkHttp}, and \textit{Elasticsearch} projects, since the \texttt{TP} metric has a risk-increasing effect in these projects (see Table~\ref{tab:model}). \textit{Spring-boot}, \textit{OkHttp}, and \textit{Elasticsearch} present \textit{positive} correlations ranging from $0.15$ (\textit{small}) up to $0.42$ (\textit{medium}). However, the \textit{Netty} project presents a \textit{medium negative} correlation of $-0.36$. Hence, despite the different behaviour in one project, we find evidence (that should be further explored in future work) indicating that the \texttt{TP} metric is a risk-increasing factor due to the complexity of developer's commits.

\textbf{\texttt{Commit Ownership}, \textit{RxJava} and \textit{Elasticsearch.}} The results of the \textbf{RQ2} show that the \texttt{CO} metric (\textbf{F6}) has a tendency to risk-decreasing effects on commit bugginess, as previously discussed~\cite{rahman2011ownership, thongtanunam2016revisiting, bird2011don}. However, the \texttt{CO} metric presents a risk-increasing effect in the \textit{RxJava} project, reaching an increase in the odds of a commit being buggy by a factor of $2.13$. This effect may be explained by a singularity in the \textit{RxJava} project. In this project, only two developers were responsible for $85.6\%$ of the bug-introducing changes reported by SZZ. Indeed, we observe that these developers are the most active ones in the project. Moreover, while the median of the \texttt{CO} metric for the remaining developers involved in the project is $0.23$, these two most active developers presented values equals to $0.70$ and $0.64$. Such fact shows that the developers responsible for the vast majority of bug-introducing changes in \textit{RxJava} work mostly on their own code. Such behavior may explain why the \texttt{CO} metric is a risk-increasing factor in the \textit{RxJava} project. A similar behavior is observed in the \textit{Elasticsearch} project, where the top 10 contributors are responsible for 75\% of the bug-introducing changes. Since the \textit{Elasticsearch} has almost ten times the size (in commits) and five times more developers than the \textit{RxJava}, we can see that the pattern holds for bigger projects. This result suggests that the risk-decreasing tendency obtained by the \texttt{CO} metric, in the vast majority of the projects, may not hold to projects where a few developers are responsible for most of the bug-introducing commits and also work mostly on their own code.

\textbf{Reengineering and Bugs.} Bavota \textit{et al.}~\cite{bavota2012does} state that several kinds of refactoring operations (\textit{e.g.}, \textit{Pull-up Method}, \textit{Inline Temp}, and many others) tend to introduce bugs very often. In our study, we also analyze the impact of refactoring operations, \textit{i.e.}, \textit{Reengineering} commits (metric \texttt{RP}), on commit bugginess. Results show that the \texttt{RP} metric presents a risk-decreasing effect in \textit{Elasticsearch} and \textit{Netty}. Such result suggests that the more focused a developer is on refactoring operations, the less likely that his commits introduce bugs in these two projects. Although our work does not deal with specific kinds of refactorings, our finding suggests that general refactoring operations decreases commit bugginess, which contradicts the finding of Bavota \textit{et al.}~\cite{bavota2012does}.


\section{Threats to Validity}
\label{sec:threats}


\textbf{Construct Validity.} The set of technical and social factors analyzed in our study may not fully represent the reasons that may lead developers to introduce bugs. To mitigate this threat, we selected factors that were analyzed by previous studies involving investigations on open-source environments~\cite{tsay2014influence, gousios2014exploratory}. We considered the perceptions of code reviewers~\cite{dabbish2012social, marlow2013impression} to define the metrics related to technical contribution norms. Nonetheless, we cannot guarantee that the community of the analyzed projects agrees with such norms. Prior work~\cite{herzig2013s} found that $33.8\%$ of the bug reports from five open-source projects were misclassified, \textit{i.e.}, a feature is requested instead of a bug reported. We mitigated this threat by performing a manual validation in all the bug reports collected. Another threat is related to correctly identify the commits that fixed bugs. To mitigate this threat, we used a GitHub functionality to identify bug-fix commits.

\textbf{Internal Validity.} We rely on the SZZ approach to locate the bug-introducing changes. Although SZZ has been widely used to locate these changes ~\cite{da2017framework}, it presents high false positive and false negative rates. To mitigate this threat, we also performed a manual validation on a sample of $922$ bug-introducing changes reported by SZZ. This validation resulted in an accuracy of $65\%$ for our SZZ implementation. However, the false negatives (\textit{i.e.}, bug-introducing changes not detected) were not included in the manual validation because of the high effort needed to validate such cases.


\textbf{External Validity.} Regarding the generality of our findings, we selected only projects in which the primary language is \textit{Java}. Although we have analyzed eight projects with different sizes, developers, and domains, our results might not hold to other projects of different primary languages.

\section{Related Work}
\label{sec:related_work}

Some previous studies focus on the relation between quality measures and commit bugginess. \'Sliwerski~\textit{et al.}~\cite{sliwerski2005changes} presented an approach to automatically locate fix-inducing commits (SZZ). They found that buggy commits are roughly three times larger than other commits. Correlations between developer characteristics (commit frequency and experience) and commit bugginess were previously investigated by Eyolfson~\textit{et al.}~\cite{eyolfson2011time}. The authors found that developers who commit on a daily basis write less buggy commits, while developers who commit as their day-job are more likely to produce bugs. Also, Eyolfson~\textit{et al.}~\cite{eyolfson2011time} suggested the existence of a correlation between developer experience and commit bugginess.

Posnett~\textit{et al.}~\cite{posnett2013dual} presented a work on ecological measures of focus in software development and their relation with bugginess. The authors defined two measures of focus (\textit{module activity focus} and \textit{developer’s attention focus}) and found that more focused developers introduce fewer bugs than less focused ones. In contrast, files that receive narrowly focused activity are more likely to contain bugs than other files.

Rahman \& Devanbu~\cite{rahman2011ownership} analyzed four open-source projects and found that high levels of ownership are associated with a lower bug introduction rate. Moreover, the authors found that specialized experience is consistently associated with buggy code, while general experience is not. Similar findings on the ownership factor were presented in the work of Bird~\textit{et al.}~\cite{bird2011don}. Thongtanunam~\textit{et al.}~\cite{thongtanunam2016revisiting} showed that there is a relationship between ownership and code review. In addition, the proportion of reviewers without expertise shares a strong relation with commit bugginess. Tufano~\textit{et al.}~\cite{tufano2017empirical} presented an empirical study on developer-related factors. Their results show that commit coherence, developer experience, and past interfering changes are associated with commit bugginess.

Mockus~\cite{mockus2010organizational} investigated the organizational factor (\textit{e.g}, size of the organization, time between releases) relating to the presence of defects in software. The author found that recent departures from an organization and distributed development are related with commit bugginess. Bernardi~\textit{et al.}~\cite{bernardi2012developers} studied the influence of developer communication on commit bugginess, finding that developers who introduce bugs have a higher social importance and communicate less. Kamei~\textit{et al.}~\cite{kamei2013large} performed a study on defect prediction for just-in-time quality assurance. The authors employed five dimensions to predict the occurrence of defects: \textit{diffusion}, \textit{size}, \textit{purpose}, \textit{history}, and \textit{experience}. Their results show that the number of files and whether or not a commit fixes a defect are related to an increase in the odds of a commit being \textit{buggy}.

Those studies~\cite{mockus2010organizational, eyolfson2011time, bernardi2012developers, bird2011don, thongtanunam2016revisiting, tufano2017empirical, kamei2013large, posnett2013dual} evaluated the relation between the factors discussed above and commit bugginess in a more limited way by considering only proprietary projects~\cite{mockus2010organizational, bird2011don}, projects that do not adopt modern code review practices~\cite{tufano2017empirical, kamei2013large} or a reduced number of factors as well as characteristics to represent them~\cite{bird2011don, thongtanunam2016revisiting, eyolfson2011time, posnett2013dual}. Our study differs from prior work by providing a more extensive study on the relation between technical, social factors, and the introduction of bugs. While prior work considers only specific factors~\cite{bird2011don, thongtanunam2016revisiting, eyolfson2011time, posnett2013dual, bernardi2012developers}, we consider a comprehensive set of multiple technical and social factors, characterized by a large set of metrics. Moreover, unlike prior work~\cite{eyolfson2011time, bird2011don, bernardi2012developers, thongtanunam2016revisiting}, we do not only analyze the effects of the metrics in isolation, but also analyze their effects grouped as factors that represent distinct aspects (technical and social) of software development. Finally, while prior work conducts their analyses at different granularity levels, such as developer~\cite{bernardi2012developers}, source file~\cite{bird2011don}, week batches~\cite{mockus2010organizational}, we conduct our analyses at the commit level, taking into account the historical aspects of software development.

\section{Conclusion}
\label{sec:conclusion}

This paper investigated the relation between different technical, social factors and the likelihood of developers to introduce bugs. We analyzed a total of $6,537$ bug reports (manually validated) and $10,674$ bug-introducing changes from eight open-source Java projects hosted on GitHub. To understand which factors may be related to the introduction of bugs, we analyzed seven different technical and social factors. First, we investigated how \textit{buggy} commits differ from \textit{clean} commits in terms of these factors. Then, we evaluated how strong is the difference between buggy and clean commits. Finally, we evaluated the effect of each factor on \textit{commit bugginess} when considering the presence of multiple factors.

Our findings show that: (i) both technical and social factors are able to discriminate between buggy and clean commits; (ii) technical factors are more determining to differentiate buggy commits from clean commits; (iii) the developers' habits of not following technical contribution norms and the developer's commit bugginess are associated with an increase on the likelihood of a new commit being buggy; and, finally, (iv) the project's establishment, the ownership level of developers' commit, and the social influence are related to a lower chance of introducing bugs. We believe that these findings benefit project managers and code reviewers, since, they may want to carefully verify contributions from developers that present factors related to commit bugginess.

As future work, 
we intend to asses the importance of contributions outside an analyzed project, to better understand developers' experience and interactions in such a complex social environment that GitHub is. Moreover, we intend to expand our work on commit bugginess to analyze a wider amount of technical (e.g., quality of the test sets, test coverage) and social (e.g., social network measures) factors.

\section*{Acknowledgments}
This work has been partially supported by CNPq (grant 427787/2018-1) and CAPES/PROCAD (grant 175956). Garcia's work has been supported by CNPq (434969/2018-4, 409536/2017-2, 312149/2016-6) and FAPERJ (22520-7/2016). Medeiro's work has been supported by CNPq (421306/2018-1, 309844/2018-5). Gheyi's work has been supported by CNPq (426005/2018-0, 308380/2016-9) and CAPES/PROCAD (117875).



\begin{thebibliography}{10}

\bibitem{herzig2013s}
Kim Herzig, Sascha Just, and Andreas Zeller.
\newblock It's not a bug, it's a feature: how misclassification impacts bug
  prediction.
\newblock In {\em Proceedings of the 2013 international conference on software
  engineering}, pages 392--401. IEEE Press, 2013.

\bibitem{Macia2012}
Isela Macia, Roberta Arcoverde, Alessandro Garcia, Christina Chavez, and Arndt
  von Staa.
\newblock {On the Relevance of Code Anomalies for Identifying Architecture
  Degradation Symptoms}.
\newblock {\em 2012 16th European Conference on Software Maintenance and
  Reengineering}, pages 277--286, March 2012.

\bibitem{marlow2013impression}
Jennifer Marlow, Laura Dabbish, and Jim Herbsleb.
\newblock Impression formation in online peer production: activity traces and
  personal profiles in github.
\newblock In {\em Proceedings of the 2013 conference on Computer supported
  cooperative work}, pages 117--128. ACM, 2013.

\bibitem{whitley2002statistics}
Elise Whitley and Jonathan Ball.
\newblock Statistics review 6: Nonparametric methods.
\newblock {\em Critical care}, 6(6):509, 2002.

\bibitem{li2006have}
Zhenmin Li, Lin Tan, Xuanhui Wang, Shan Lu, Yuanyuan Zhou, and Chengxiang Zhai.
\newblock Have things changed now?: an empirical study of bug characteristics
  in modern open source software.
\newblock In {\em Proceedings of the 1st workshop on Architectural and system
  support for improving software dependability}, pages 25--33. ACM, 2006.

\bibitem{wohlin2012experimentation}
Claes Wohlin, Per Runeson, Martin H{\"o}st, Magnus~C Ohlsson, Bj{\"o}rn
  Regnell, and Anders Wessl{\'e}n.
\newblock {\em Experimentation in software engineering}.
\newblock Springer Science \& Business Media, 2012.

\bibitem{williams2008szz}
Chadd Williams and Jaime Spacco.
\newblock Szz revisited: verifying when changes induce fixes.
\newblock In {\em Proceedings of the 2008 workshop on Defects in large software
  systems}, pages 32--36. ACM, 2008.

\bibitem{kim2006automatic}
Sunghun Kim, Thomas Zimmermann, Kai Pan, E~James~Jr, et~al.
\newblock Automatic identification of bug-introducing changes.
\newblock In {\em Automated Software Engineering, 2006. ASE'06. 21st IEEE/ACM
  International Conference on}, pages 81--90. IEEE, 2006.

\bibitem{gala2013intensive}
Santiago Gala-P{\'e}rez, Gregorio Robles, Jes{\'u}s~M Gonz{\'a}lez-Barahona,
  and Israel Herraiz.
\newblock Intensive metrics for the study of the evolution of open source
  projects: Case studies from apache software foundation projects.
\newblock In {\em Proceedings of the 10th Working Conference on Mining Software
  Repositories}, pages 159--168. IEEE Press, 2013.

\bibitem{gousios2014exploratory}
Georgios Gousios, Martin Pinzger, and Arie~van Deursen.
\newblock An exploratory study of the pull-based software development model.
\newblock In {\em Proceedings of the 36th International Conference on Software
  Engineering}, pages 345--355. ACM, 2014.

\bibitem{blincoe2015ecosystems}
Kelly Blincoe, Francis Harrison, and Daniela Damian.
\newblock Ecosystems in github and a method for ecosystem identification using
  reference coupling.
\newblock In {\em Proceedings of the 12th Working Conference on Mining Software
  Repositories}, pages 202--207. IEEE Press, 2015.

\bibitem{gousios2016work}
Georgios Gousios, Margaret-Anne Storey, and Alberto Bacchelli.
\newblock Work practices and challenges in pull-based development: the
  contributor's perspective.
\newblock In {\em Software Engineering (ICSE), 2016 IEEE/ACM 38th International
  Conference on}, pages 285--296. IEEE, 2016.

\bibitem{yu2015wait}
Yue Yu, Huaimin Wang, Vladimir Filkov, Premkumar Devanbu, and Bogdan Vasilescu.
\newblock Wait for it: Determinants of pull request evaluation latency on
  github.
\newblock In {\em Mining software repositories (MSR), 2015 IEEE/ACM 12th
  working conference on}, pages 367--371. IEEE, 2015.

\bibitem{vasilescu2014continuous}
Bogdan Vasilescu, Stef Van~Schuylenburg, Jules Wulms, Alexander Serebrenik, and
  Mark~GJ van~den Brand.
\newblock Continuous integration in a social-coding world: Empirical evidence
  from github.
\newblock In {\em Software Maintenance and Evolution (ICSME), 2014 IEEE
  International Conference on}, pages 401--405. IEEE, 2014.

\bibitem{kalliamvakou2014promises}
Eirini Kalliamvakou, Georgios Gousios, Kelly Blincoe, Leif Singer, Daniel~M
  German, and Daniela Damian.
\newblock The promises and perils of mining github.
\newblock In {\em Proceedings of the 11th working conference on mining software
  repositories}, pages 92--101. ACM, 2014.

\bibitem{dabbish2012social}
Laura Dabbish, Colleen Stuart, Jason Tsay, and Jim Herbsleb.
\newblock Social coding in github: transparency and collaboration in an open
  software repository.
\newblock In {\em Proceedings of the ACM 2012 conference on Computer Supported
  Cooperative Work}, pages 1277--1286. ACM, 2012.

\bibitem{sliwerski2005changes}
Jacek {\'S}liwerski, Thomas Zimmermann, and Andreas Zeller.
\newblock When do changes induce fixes?
\newblock In {\em ACM sigsoft software engineering notes}, volume~30, pages
  1--5. ACM, 2005.

\bibitem{tsay2014influence}
Jason Tsay, Laura Dabbish, and James Herbsleb.
\newblock Influence of social and technical factors for evaluating contribution
  in github.
\newblock pages 356--366. ACM, 2014.

\bibitem{mukaka2012guide}
Mavuto~M Mukaka.
\newblock A guide to appropriate use of correlation coefficient in medical
  research.
\newblock {\em Malawi Medical Journal}, 24(3):69--71, 2012.

\bibitem{eyolfson2011time}
Jon Eyolfson, Lin Tan, and Patrick Lam.
\newblock Do time of day and developer experience affect commit bugginess?
\newblock In {\em Proceedings of the 8th Working Conference on Mining Software
  Repositories}, pages 153--162. ACM, 2011.

\bibitem{soto2017analyzing}
Mauricio Soto, Zack Coker, and Claire Le~Goues.
\newblock Analyzing the impact of social attributes on commit integration
  success.
\newblock In {\em Mining Software Repositories (MSR), 2017 IEEE/ACM 14th
  International Conference on}, pages 483--486. IEEE, 2017.

\bibitem{terrell2017gender}
Josh Terrell, Andrew Kofink, Justin Middleton, Clarissa Rainear, Emerson
  Murphy-Hill, Chris Parnin, and Jon Stallings.
\newblock Gender differences and bias in open source: Pull request acceptance
  of women versus men.
\newblock {\em PeerJ Computer Science}, 3:e111, 2017.

\bibitem{hattori2008nature}
Lile~P Hattori and Michele Lanza.
\newblock On the nature of commits.
\newblock In {\em Proceedings of the 23rd IEEE/ACM International Conference on
  Automated Software Engineering}, pages III--63. IEEE Press, 2008.

\bibitem{hindle2008large}
Abram Hindle, Daniel~M German, and Ric Holt.
\newblock What do large commits tell us?: a taxonomical study of large commits.
\newblock In {\em Proceedings of the 2008 international working conference on
  Mining software repositories}, pages 99--108. ACM, 2008.

\bibitem{purushothaman2005toward}
Ranjith Purushothaman and Dewayne~E Perry.
\newblock Toward understanding the rhetoric of small source code changes.
\newblock {\em IEEE Transactions on Software Engineering}, 31(6):511--526,
  2005.

\bibitem{mockus2000identifying}
Audris Mockus and Lawrence~G. Votta.
\newblock Identifying reasons for software changes using historic databases.
\newblock In {\em Proceedings of the International Conference on Software
  Maintenance (ICSM'00)}, pages 120--. IEEE Computer Society, 2000.

\bibitem{dragan2011using}
Natalia Dragan, Michael~L Collard, Maen Hammad, and Jonathan~I Maletic.
\newblock Using stereotypes to help characterize commits.
\newblock In {\em Software Maintenance (ICSM), 2011 27th IEEE International
  Conference on}, pages 520--523. IEEE, 2011.

\bibitem{mockus2010organizational}
Audris Mockus.
\newblock Organizational volatility and its effects on software defects.
\newblock In {\em Proceedings of the eighteenth ACM SIGSOFT international
  symposium on Foundations of software engineering}, pages 117--126. ACM, 2010.

\bibitem{rahman2011ownership}
Foyzur Rahman and Premkumar Devanbu.
\newblock Ownership, experience and defects: a fine-grained study of
  authorship.
\newblock In {\em Proceedings of the 33rd International Conference on Software
  Engineering}, pages 491--500. ACM, 2011.

\bibitem{kamei2013large}
Yasutaka Kamei, Emad Shihab, Bram Adams, Ahmed~E Hassan, Audris Mockus, Anand
  Sinha, and Naoyasu Ubayashi.
\newblock A large-scale empirical study of just-in-time quality assurance.
\newblock {\em IEEE Transactions on Software Engineering}, 39(6):757--773,
  2013.

\bibitem{bacchelli2013expectations}
Alberto Bacchelli and Christian Bird.
\newblock Expectations, outcomes, and challenges of modern code review.
\newblock In {\em Proceedings of the 2013 international conference on software
  engineering}, pages 712--721. IEEE Press, 2013.

\bibitem{thongtanunam2016revisiting}
Patanamon Thongtanunam, Shane McIntosh, Ahmed~E Hassan, and Hajimu Iida.
\newblock Revisiting code ownership and its relationship with software quality
  in the scope of modern code review.
\newblock In {\em Proceedings of the 38th international conference on software
  engineering}, pages 1039--1050. ACM, 2016.

\bibitem{bernardi2012developers}
Mario~Luca Bernardi, Gerardo Canfora, Giuseppe~A Di~Lucca, Massimiliano
  Di~Penta, and Damiano Distante.
\newblock Do developers introduce bugs when they do not communicate? the case
  of eclipse and mozilla.
\newblock In {\em Software Maintenance and Reengineering (CSMR), 2012 16th
  European Conference on}, pages 139--148. IEEE, 2012.

\bibitem{bird2011don}
Christian Bird, Nachiappan Nagappan, Brendan Murphy, Harald Gall, and Premkumar
  Devanbu.
\newblock Don't touch my code!: examining the effects of ownership on software
  quality.
\newblock In {\em Proceedings of the 19th ACM SIGSOFT symposium and the 13th
  European conference on Foundations of software engineering}, pages 4--14.
  ACM, 2011.

\bibitem{da2017framework}
Daniel~Alencar da~Costa, Shane McIntosh, Weiyi Shang, Uir{\'a} Kulesza, Roberta
  Coelho, and Ahmed~E Hassan.
\newblock A framework for evaluating the results of the szz approach for
  identifying bug-introducing changes.
\newblock {\em IEEE Transactions on Software Engineering}, 43(7):641--657,
  2017.

\bibitem{kalliamvakou2016depth}
Eirini Kalliamvakou, Georgios Gousios, Kelly Blincoe, Leif Singer, Daniel~M
  German, and Daniela Damian.
\newblock An in-depth study of the promises and perils of mining github.
\newblock {\em Empirical Software Engineering}, 21(5):2035--2071, 2016.

\bibitem{caretpackage}
Max Kuhn.
\newblock Building predictive models in r using the caret package.
\newblock {\em Journal of Statistical Software, Articles}, 28(5):1--26, 2008.

\bibitem{dormann2013collinearity}
Carsten~F Dormann, Jane Elith, Sven Bacher, Carsten Buchmann, Gudrun Carl,
  Gabriel Carr{\'e}, Jaime R~Garc{\'\i}a Marqu{\'e}z, Bernd Gruber, Bruno
  Lafourcade, Pedro~J Leit{\~a}o, et~al.
\newblock Collinearity: a review of methods to deal with it and a simulation
  study evaluating their performance.
\newblock {\em Ecography}, 36(1):27--46, 2013.

\bibitem{rlang}
{R Core Team}.
\newblock {\em R: A Language and Environment for Statistical Computing}.
\newblock R Foundation for Statistical Computing, 2018.

\bibitem{guisan2000predictive}
Antoine Guisan and Niklaus~E Zimmermann.
\newblock Predictive habitat distribution models in ecology.
\newblock {\em Ecological modelling}, 135(2-3):147--186, 2000.

\bibitem{zhou2010developer}
Minghui Zhou and Audris Mockus.
\newblock Developer fluency: Achieving true mastery in software projects.
\newblock In {\em Proceedings of the eighteenth ACM SIGSOFT international
  symposium on Foundations of software engineering}, pages 137--146. ACM, 2010.

\bibitem{guo2010characterizing}
Philip~J Guo, Thomas Zimmermann, Nachiappan Nagappan, and Brendan Murphy.
\newblock Characterizing and predicting which bugs get fixed: an empirical
  study of microsoft windows.
\newblock In {\em Software Engineering, 2010 ACM/IEEE 32nd International
  Conference on}, volume~1, pages 495--504. IEEE, 2010.

\bibitem{kim2008classifying}
Sunghun Kim, E~James Whitehead~Jr, and Yi~Zhang.
\newblock Classifying software changes: Clean or buggy?
\newblock {\em IEEE Transactions on Software Engineering}, 34(2):181--196,
  2008.

\bibitem{nakakoji2002evolution}
Kumiyo Nakakoji, Yasuhiro Yamamoto, Yoshiyuki Nishinaka, Kouichi Kishida, and
  Yunwen Ye.
\newblock Evolution patterns of open-source software systems and communities.
\newblock In {\em Proceedings of the international workshop on Principles of
  software evolution}, pages 76--85. ACM, 2002.

\bibitem{posnett2013dual}
Daryl Posnett, Raissa D'Souza, Premkumar Devanbu, and Vladimir Filkov.
\newblock Dual ecological measures of focus in software development.
\newblock In {\em Software Engineering (ICSE), 2013 35th International
  Conference on}, pages 452--461. IEEE, 2013.

\bibitem{grissom2005effect}
Robert~J Grissom and John~J Kim.
\newblock {\em Effect sizes for research: A broad practical approach.}
\newblock Lawrence Erlbaum Associates Publishers, 2005.

\bibitem{romano2006exploring}
Jeanine Romano, Jeffrey~D Kromrey, Jesse Coraggio, Jeff Skowronek, and Linda
  Devine.
\newblock Exploring methods for evaluating group differences on the nsse and
  other surveys: Are the t-test and cohen’sd indices the most appropriate
  choices.
\newblock pages 1--51. Citeseer, 2006.

\bibitem{effsizepackage}
Marco Torchiano.
\newblock {\em effsize: Efficient Effect Size Computation}, 2017.
\newblock R package version 0.7.1.

\bibitem{tufano2017empirical}
Michele Tufano, Gabriele Bavota, Denys Poshyvanyk, Massimiliano Di~Penta, Rocco
  Oliveto, and Andrea De~Lucia.
\newblock An empirical study on developer-related factors characterizing
  fix-inducing commits.
\newblock {\em Journal of Software: Evolution and Process}, 29(1):e1797, 2017.

\bibitem{sullivan2012using}
Gail~M Sullivan and Richard Feinn.
\newblock Using effect size—or why the p value is not enough.
\newblock {\em Journal of graduate medical education}, 4(3):279--282, 2012.

\bibitem{bavota2012does}
Gabriele Bavota, Bernardino De~Carluccio, Andrea De~Lucia, Massimiliano
  Di~Penta, Rocco Oliveto, and Orazio Strollo.
\newblock When does a refactoring induce bugs? an empirical study.
\newblock In {\em Source Code Analysis and Manipulation (SCAM), 2012 IEEE 12th
  International Working Conference on}, pages 104--113. IEEE, 2012.

\bibitem{cohen2013statistical}
Jacob Cohen.
\newblock {\em Statistical power analysis for the behavioral sciences}.
\newblock Academic Press, 2013.

\bibitem{mcdonald2009handbook}
John~H McDonald.
\newblock {\em Handbook of biological statistics}, volume~2.
\newblock Sparky House Publishing, 2009.

\bibitem{steinmacher2018almost}
Igor Steinmacher, Gustavo Pinto, Igor~Scaliante Wiese, and Marco~Aur{\'e}lio
  Gerosa.
\newblock Almost there: A study on quasi-contributors in open-source software
  projects.
\newblock In {\em 2018 IEEE/ACM 40th International Conference on Software
  Engineering (ICSE)}, pages 256--266. IEEE, 2018.

\bibitem{d2012evaluating}
Marco D’Ambros, Michele Lanza, and Romain Robbes.
\newblock Evaluating defect prediction approaches: a benchmark and an extensive
  comparison.
\newblock {\em Empirical Software Engineering}, 17(4-5):531--577, 2012.

\bibitem{moser2008comparative}
Raimund Moser, Witold Pedrycz, and Giancarlo Succi.
\newblock A comparative analysis of the efficiency of change metrics and static
  code attributes for defect prediction.
\newblock In {\em Proceedings of the 30th international conference on Software
  engineering}, pages 181--190. ACM, 2008.

\bibitem{Falcao2019}
Filipe Falcão, Caio Barbosa, Baldoino Fonseca, Alessandro Garcia, Márcio
  Ribeiro, França Sales, Rohit Gheyi, and Elder Cirilo.
\newblock Assessing the influence of developers' practices on the introduction
  of bugs (under review).
\newblock {\em Software Quality Journal}, 2019.

\end{thebibliography}


\begin{thebibliography}{00}
\IEEEtriggeratref{18}
\bibitem{tsay2014influence}
Jason Tsay, Laura Dabbish, and James Herbsleb.
\newblock Influence of social and technical factors for evaluating contribution in GitHub.
\newblock In {\em Proceedings of the 36th International Conference on Software
  Engineering (ICSE)}, pages 356--366. ACM, 2014.

\bibitem{sliwerski2005changes}
Jacek {\'S}liwerski, Thomas Zimmermann, and Andreas Zeller.
\newblock When do changes induce fixes?
\newblock In {\em ACM sigsoft software engineering notes}, volume~30, pages
  1--5. ACM, 2005.

\bibitem{gousios2014exploratory}
Georgios Gousios, Martin Pinzger, and Arie~van Deursen.
\newblock An exploratory study of the pull-based software development model.
\newblock In {\em Proceedings of the 36th International Conference on Software
  Engineering (ICSE)}, pages 345--355. ACM, 2014.

\bibitem{gousios2016work}
Georgios Gousios, Margaret-Anne Storey, and Alberto Bacchelli.
\newblock Work practices and challenges in pull-based development: the
  contributor's perspective.
\newblock In {\em Proceedings of the 38th International
  Conference on Software Engineering (ICSE)}, pages 285--296. IEEE, 2016.

\bibitem{eyolfson2011time}
Jon Eyolfson, Lin Tan, and Patrick Lam.
\newblock Do time of day and developer experience affect commit bugginess?
\newblock In {\em Proceedings of the 8th Working Conference on Mining Software
  Repositories (MSR)}, pages 153--162. ACM, 2011.

\bibitem{mockus2010organizational}
Audris Mockus.
\newblock Organizational volatility and its effects on software defects.
\newblock In {\em Proceedings of the Foundations of Software Engineering (FSE)}, pages     117--126. ACM, 2010.

\bibitem{bernardi2012developers}
Mario~Luca Bernardi, Gerardo Canfora, Giuseppe~A Di~Lucca, Massimiliano
  Di~Penta, and Damiano Distante.
\newblock Do developers introduce bugs when they do not communicate? the case
  of eclipse and mozilla.
\newblock In {\em Proceedings of the 16th
  European Conference on Software Maintenance and Reengineering (CSMR)}, pages 139--148. IEEE, 2012.

\bibitem{bird2011don}
Christian Bird, Nachiappan Nagappan, Brendan Murphy, Harald Gall, and Premkumar
  Devanbu.
\newblock Don't touch my code!: examining the effects of ownership on software
  quality.
\newblock In {\em Proceedings of the Foundations of Software Engineering (FSE)}, pages 4--14.
  ACM, 2011.

\bibitem{guo2010characterizing}
Philip~J Guo, Thomas Zimmermann, Nachiappan Nagappan, and Brendan Murphy.
\newblock Characterizing and predicting which bugs get fixed: an empirical
  study of microsoft windows.
\newblock In {\em Proceedings of the 32th International Conference on Software Engineering (ICSE)}, volume~1, pages 495--504. IEEE, 2010.

\bibitem{posnett2013dual}
Daryl Posnett, Raissa D'Souza, Premkumar Devanbu, and Vladimir Filkov.
\newblock Dual ecological measures of focus in software development.
\newblock In {\em Proceedings of the 35th International Conference on Software Engineering (ICSE)}, pages 452--461. IEEE, 2013.

\bibitem{kamei2013large}
Yasutaka Kamei, Emad Shihab, Bram Adams, Ahmed~E Hassan, Audris Mockus, Anand
  Sinha, and Naoyasu Ubayashi.
\newblock A large-scale empirical study of just-in-time quality assurance.
\newblock {\em IEEE Transactions on Software Engineering}, 39(6):757--773,
  2013.

\bibitem{thongtanunam2016revisiting}
Patanamon Thongtanunam, Shane McIntosh, Ahmed~E Hassan, and Hajimu Iida.
\newblock Revisiting code ownership and its relationship with software quality
  in the scope of modern code review.
\newblock In {\em Proceedings of the 38th International Conference on Software
  Engineering (ICSE)}, pages 1039--1050. ACM, 2016.

\bibitem{tufano2017empirical}
Michele Tufano, Gabriele Bavota, Denys Poshyvanyk, Massimiliano Di~Penta, Rocco
  Oliveto, and Andrea De~Lucia.
\newblock An empirical study on developer-related factors characterizing
  fix-inducing commits.
\newblock {\em Journal of Software: Evolution and Process}, 29(1):e1797, 2017.

\bibitem{rahman2011ownership}
Foyzur Rahman and Premkumar Devanbu.
\newblock Ownership, experience and defects: a fine-grained study of
  authorship.
\newblock In {\em Proceedings of the 33rd International Conference on Software
  Engineering (ICSE)}, pages 491--500. ACM, 2011.

\bibitem{hattori2008nature}
Lile~P Hattori and Michele Lanza.
\newblock On the nature of commits.
\newblock In {\em Proceedings of the 23rd International Conference on
  Automated Software Engineering (ASE)}, pages 63--71. IEEE Press, 2008.

\bibitem{mockus2000identifying}
Audris Mockus and Lawrence~G. Votta.
\newblock Identifying reasons for software changes using historic databases.
\newblock In {\em Proceedings of the International Conference on Software
  Maintenance (ICSM)}, pages 120--. IEEE Computer Society, 2000.

\bibitem{dragan2011using}
Natalia Dragan, Michael~L Collard, Maen Hammad, and Jonathan~I Maletic.
\newblock Using stereotypes to help characterize commits.
\newblock In {\em Proceedings of the International
  Conference on Software Maintenance (ICSM)}, pages 520--523. IEEE, 2011.

\bibitem{dabbish2012social}
Laura Dabbish, Colleen Stuart, Jason Tsay, and Jim Herbsleb.
\newblock Social coding in GitHub: transparency and collaboration in an open
  software repository.
\newblock In {\em Proceedings of the ACM 2012 conference on Computer Supported
  Cooperative Work}, pages 1277--1286. ACM, 2012.

\bibitem{purushothaman2005toward}
Ranjith Purushothaman and Dewayne~E Perry.
\newblock Toward understanding the rhetoric of small source code changes.
\newblock {\em IEEE Transactions on Software Engineering}, 31(6):511--526,
  2005.

\bibitem{nakakoji2002evolution}
Kumiyo Nakakoji, Yasuhiro Yamamoto, Yoshiyuki Nishinaka, Kouichi Kishida, and
  Yunwen Ye.
\newblock Evolution patterns of open-source software systems and communities.
\newblock In {\em Proceedings of the international workshop on Principles of
  software evolution}, pages 76--85. ACM, 2002.

\bibitem{herzig2013s}
Kim Herzig, Sascha Just, and Andreas Zeller.
\newblock It's not a bug, it's a feature: how misclassification impacts bug
  prediction.
\newblock In {\em Proceedings of the International Conference on Software
  Engineering (ICSE)}, pages 392--401. IEEE Press, 2013.

\bibitem{marlow2013impression}
Jennifer Marlow, Laura Dabbish, and Jim Herbsleb.
\newblock Impression formation in online peer production: activity traces and
  personal profiles in GitHub.
\newblock In {\em Proceedings of the 2013 conference on Computer supported
  cooperative work}, pages 117--128. ACM, 2013.

\bibitem{whitley2002statistics}
Elise Whitley and Jonathan Ball.
\newblock Statistics review 6: Nonparametric methods.
\newblock {\em Critical care}, 6(6):509, 2002.

\bibitem{wohlin2012experimentation}
Claes Wohlin, Per Runeson, Martin H{\"o}st, Magnus~C Ohlsson, Bj{\"o}rn
  Regnell, and Anders Wessl{\'e}n.
\newblock {\em Experimentation in software engineering}.
\newblock Springer Science \& Business Media, 2012.

\bibitem{williams2008szz}
Chadd Williams and Jaime Spacco.
\newblock SZZ revisited: verifying when changes induce fixes.
\newblock In {\em Proceedings of the 2008 workshop on Defects in large software
  systems}, pages 32--36. ACM, 2008.

\bibitem{kim2006automatic}
Sunghun Kim, Thomas Zimmermann, Kai Pan, E~James~Jr, et~al.
\newblock Automatic identification of bug-introducing changes.
\newblock In {\em Proceedings of the 21st International Conference on Automated Software Engineering (ASE)}, pages 81--90. IEEE, 2006.

\bibitem{bacchelli2013expectations}
Alberto Bacchelli and Christian Bird.
\newblock Expectations, outcomes, and challenges of modern code review.
\newblock In {\em Proceedings of the 2013 International Conference on Software
  Engineering (ICSE)}, pages 712--721. IEEE Press, 2013.

\bibitem{da2017framework}
Daniel~Alencar da~Costa, Shane McIntosh, Weiyi Shang, Uir{\'a} Kulesza, Roberta
  Coelho, and Ahmed~E Hassan.
\newblock A framework for evaluating the results of the SZZ approach for
  identifying bug-introducing changes.
\newblock {\em IEEE Transactions on Software Engineering}, 43(7):641--657,
  2017.

\bibitem{dormann2013collinearity}
Carsten~F Dormann, Jane Elith, Sven Bacher, Carsten Buchmann, Gudrun Carl,
  Gabriel Carr{\'e}, Jaime R~Garc{\'\i}a Marqu{\'e}z, Bernd Gruber, Bruno
  Lafourcade, Pedro~J Leit{\~a}o, et~al.
\newblock Collinearity: a review of methods to deal with it and a simulation
  study evaluating their performance.
\newblock {\em Ecography}, 36(1):27--46, 2013.

\bibitem{grissom2005effect}
Robert~J Grissom and John~J Kim.
\newblock {\em Effect sizes for research: A broad practical approach.}
\newblock Lawrence Erlbaum Associates Publishers, 2005.

\bibitem{romano2006exploring}
Jeanine Romano, Jeffrey~D Kromrey, Jesse Coraggio, Jeff Skowronek, and Linda
  Devine.
\newblock Exploring methods for evaluating group differences on the nsse and
  other surveys: Are the t-test and cohen’sd indices the most appropriate
  choices.
\newblock pages 1--51. Citeseer, 2006.

\bibitem{sullivan2012using}
Gail~M Sullivan and Richard Feinn.
\newblock Using effect size—or why the p value is not enough.
\newblock {\em Journal of graduate medical education}, 4(3):279--282, 2012.

\bibitem{bavota2012does}
Gabriele Bavota, Bernardino De~Carluccio, Andrea De~Lucia, Massimiliano
  Di~Penta, Rocco Oliveto, and Orazio Strollo.
\newblock When does a refactoring induce bugs? an empirical study.
\newblock In {\em Proceedings of the 12th
  International Working Conference on Source Code Analysis and Manipulation (SCAM)}, pages 104--113. IEEE, 2012.

\bibitem{cohen2013statistical}
Jacob Cohen.
\newblock {\em Statistical power analysis for the behavioral sciences}.
\newblock Academic Press, 2013.

\bibitem{mcdonald2009handbook}
John~H McDonald.
\newblock {\em Handbook of biological statistics}, volume~2.
\newblock Sparky House Publishing, 2009.

\bibitem{steinmacher2018almost}
Igor Steinmacher, Gustavo Pinto, Igor Scaliante Wiese, and Marco~Aur{\'e}lio
  Gerosa.
\newblock Almost there: A study on quasi-contributors in open-source software
  projects.
\newblock In {\em Proceedings of the 40th International Conference on Software
  Engineering (ICSE)}, pages 256--266. IEEE, 2018.

\end{thebibliography}
\end{document}